\documentclass[twocolumn,showpacs,showkeys]{revtex4}
\usepackage{graphicx}

\newcommand*{\rav}{\ensuremath{R_\mathrm{av}}}
\newcommand*{\pw}{PW/cm$^2$}  

\begin{document}


\title{Kinetic enhancement of Raman backscatter, and electron acoustic Thomson scatter}
\date{\today}
\author{D.\ J.\ Strozzi}
\email{dstrozzi@llnl.gov}
\author{E.\ A.\ Williams}
\author{A.\ B.\ Langdon}
\affiliation{Lawrence Livermore National Laboratory, University of California, Livermore, CA 94550}
\author{A.\ Bers}
\affiliation{Massachusetts Institute of Technology, Cambridge, MA 02139}

\begin{abstract}
1-D Eulerian Vlasov-Maxwell simulations are presented which show kinetic enhancement of stimulated Raman backscatter (SRBS) due to electron trapping in regimes of heavy linear Landau damping. The conventional Raman Langmuir wave is transformed into a set of beam acoustic modes [L.\ Yin \textit{et al.}, Phys.\ Rev.\ E \textbf{73}, 025401 (2006)]. For the first time, a low phase velocity electron acoustic wave (EAW) is seen developing from the self-consistent Raman physics. Backscatter of the pump laser off the EAW fluctuations is reported and referred to as electron acoustic Thomson scatter. This light is similar in wavelength to, although much lower in amplitude than, the reflected light between the pump and SRBS wavelengths observed in single hot spot experiments, and previously interpreted as stimulated electron acoustic scatter [D.\ S.\ Montgomery \textit{et al.}, Phys.\ Rev.\ Lett.\ \textbf{87}, 155001 (2001)]. The EAW is strongest well below the phase-matched frequency for electron acoustic scatter, and therefore the EAW is not produced by it. The beating of different beam acoustic modes is proposed as the EAW excitation mechanism, and is called beam acoustic decay. Supporting evidence for this process, including bispectral analysis, is presented. The linear electrostatic modes, found by projecting the numerical distribution function onto a Gauss-Hermite basis, include beam acoustic modes (some of which are unstable even without parametric coupling to light waves) and a strongly-damped EAW similar to the observed one. This linear EAW results from non-Maxwellian features in the electron distribution, rather than nonlinearity due to electron trapping.
\end{abstract}


\maketitle

\section{Introduction}
The role of kinetic effects in stimulated Raman backscatter (SRBS) is of much current interest. SRBS is the three-wave parametric coupling of a pump light wave (the laser, labeled mode 0) to a counter-propagating daughter light wave (mode 1) and co-propagating electron plasma wave (EPW, mode 2), and satisfies the resonance conditions $\vec{k}_0=\vec{k}_1+\vec{k}_2$ and $\omega_0=\omega_1+\omega_2$. It may remove a substantial amount of energy from a high-intensity laser propagating through a plasma and create energetic electrons \cite{kruer-lpi-1988}. SRBS and other laser-plasma interactions must be sufficiently controlled for laser-driven inertial fusion to succeed. In the high-temperature plasmas expected on ignition experiments such as the National Ignition Facility (NIF) \cite{lindl-nif-pop-2004} and Laser M\'egajoule (LMJ) \cite{cavailler-lmj-ppcf-2005}, linear theory of parametric instabilities predicts SRBS to have a weak spatial gain rate due to the large $k_2\lambda_D$ ($\lambda_D \equiv [\epsilon_0T_e/(n_0e^2)]^{1/2}$ is the electron Debye length) and resulting heavy Landau damping. However, recent experiments \cite{fernandez-srs-pop-2000, montgomery-seas-prl-2001, montgomery-trident-pop-2002} and kinetic simulations \cite{vu-kininf-prl-2001, vu-kininf-pop-2002, yin-sbams-pre-2006, yin-sbams-pop-2006, strozzi-thesis-2005, strozzi-srs-jpp-2005} have shown SRBS reflectivities well above coupled-mode convective gain levels.

This kinetic inflation or enhancement of SRBS is widely attributed to electron trapping in the EPW potential well, which flattens the electron distribution function near the EPW phase velocity $v_{p2}$ \cite{vu-kininf-prl-2001,kline-lwnonlin-pop-2006}. As has long been known, trapping reduces the EPW damping rate \cite{oneil-damping-pof-1965} and induces an amplitude-dependent frequency downshift \cite{morales-freqshift-prl-1972}. SRBS saturation results from the competition of nonlinear effects including the damping reduction, frequency shift, pump depletion, and trapped particle instability \cite{brunner-valeo-prl-2004}, and possibly spatio-temporal chaos after ion modes build up near the laser entrance \cite{salcedo-srs-nf-2003}, among others. A theory for the onset of kinetic inflation and the resulting time-averaged reflectivity is not yet in hand, but would be of great value for designing laser fusion implosions.

This current work is aimed at better understanding kinetic effects such as electron trapping in SRBS for time scales and frequencies such that ion dynamics can be ignored. To this end, we have performed 1-D Eulerian Vlasov-Maxwell simulations of SRBS using the ELVIS code \cite{strozzi-srs-cpc-2004, strozzi-srs-jpp-2005, strozzi-thesis-2005}. A rapid increase of reflectivity above the convective value is seen as we vary parameters such as the pump intensity. The system does not reach a temporal steady state when kinetic enhancement occurs, but instead the reflectivity comes in sub-picosecond bursts (as seen in, e.g., Ref. \cite{langdon-rescat-prl-2002}). Besides enhancing SRBS, trapping also changes the electrostatic modes present. Electrostatic spectra from our simulations reveal the EPW dispersion curve splits into an upper branch and several beam acoustic modes (BAMs) \cite{yin-sbams-pre-2006, yin-sbams-pop-2006}, and that SRBS involves one of the latter. The SRBS daughter BAM activity occurs along the streak that phase-matches for pump decay, which extends down in frequency from the linear phase-matching point with a slope equal to the daughter light wave's group velocity $v_{g1}\approx-c$. This behavior has been seen in Thomson scattering spectra from recent Trident experiments \cite{kline-srs-prl-2005} and PIC simulations \cite{yin-sbams-pre-2006}.

A low-amplitude acoustic wave (meaning that $\omega\propto k$, not an ion acoustic wave), which we call the electron acoustic wave (EAW), also appears once inflation develops in our simulations. Moreover, weak reflected light that phase-matches for scattering off this mode is also seen, which we call electron acoustic scatter (EAS). This process is reminiscent of what was called stimulated electron acoustic scatter (SEAS) and observed in Trident single hot spot experiments \cite{ montgomery-seas-prl-2001, montgomery-trident-pop-2002} (although at the most intense pumps, the experimental SEAS levels relative to SRBS are much stronger than in our simulations). To our knowledge, this paper is the first to analyze in detail EAWs and EAS arising from the SRBS dynamics, and not from manually-distorted distributions or other strong seeding. SEAS has also been reported in PIC simulations of plasmas overdense to SRBS and at relativistic pump intensities \cite{nikolic-seas-pre-2002}, and may have been seen in underdense Vlasov simulations \cite{sircombe-eaw-ppcf-2006}.

Most of the EAW energy in our simulations is much too low-frequency to match EAS and must be produced independently. We propose the three-wave coupling of two BAMs and an EAW, which can be called beam acoustic decay (BAD), as the EAW excitation mechanism. Smaller-amplitude EAWs at higher frequency are subsequently generated by harmonic coupling. The pump laser then scatters off these fluctuations, referred to below as electron acoustic Thomson scatter (EATS). The phase-matching needed for BAD and EATS is seen both in electrostatic $(k,\omega)$ spectra and bispectral analysis.

We study the electrostatic mode structure by solving the linear dispersion relation for simulation distributions projected onto a Gauss-Hermite basis. This yields the BAMs and EAW seen in our simulation spectra, and shows a damping reduction and frequency downshift (compared to the EPW for a Maxwellian) of the BAM involved in SRBS. The BAMs are linearly unstable - in the absence of parametric coupling to light waves - for some wavenumber range. In addition, we find an EAW root with heavy damping. This linear EAW differs from the nonlinear EAW due to trapping discussed by others \cite{holloway-epws-pra-1991, schamel-holeq-pop-2000, rose-nonlinEPW-pop-2001, rose-trappedSRS-pop-2003, valentini-eaw-pop-2006}, which unlike ours is undamped and cuts off at $k\lambda_D\approx 0.53$ (in the limit of zero field amplitude).

The paper is organized as follows: Section \ref{sec:model} presents the model equations and geometry used in the ELVIS code. The dependence of SRBS on pump strength for simulations of Trident single hot spot conditions is discussed in Sec.\ \ref{sec:tridgen}.  Kinetically-inflated SRBS, and the related beam acoustic modes and electron acoustic scatter, are carefully studied in Sec.\ \ref{sec:trid2e15} for a pump intensity of 2 \pw . Section \ref{sec:hohl} demonstrates that similar physics occurs in higher temperature, higher density plasmas found in hohlraum fills. We discuss our results (including experimental relevance) and conclude in Sec.\ \ref{sec:conc}. The Appendix details the Gauss-Hermite mode-finding method.

\section{Model and Numerical Approach}
\label{sec:model}

\begin{figure}
  \includegraphics{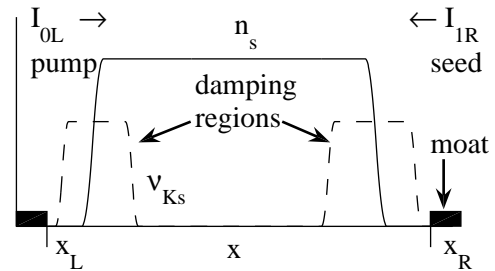}
  \caption{\label{fig:model_geom} 1-D geometry of the ELVIS code.}
\end{figure}

We use the Eulerian Vlasov-Maxwell solver ELVIS \cite{strozzi-srs-cpc-2004, strozzi-srs-jpp-2005, strozzi-thesis-2005}, which resembles the code presented in \cite{ghizzo-vlasov-jcp-1990}. Our code implements the simplest model which allows for electron kinetic effects in Raman scattering. The geometry is one-dimensional, with all spatial variation and wavevectors in the $x$ direction. We describe the particle species by the nonrelativistic Vlasov equation in $x$ and keep the ions fixed in this paper. The light waves are linearly polarized in $y$, and the particles constitute a cold collisionless fluid in this direction (the light waves are undamped). The geometry is illustrated in Fig.~\ref{fig:model_geom}.

The governing equations are:
 \begin{eqnarray}
 \Big[ \partial_{t} + v_{x}\partial_{x} &+& \left( Z_{s}e/m_s\right)
   \left( E_{x}+v_{ys}B_{z}\right) \partial _{v_{x}} \Big] f_{s} \\
   &=& \nu_{Ks}\left( x\right) \left( n_{s}\hat{f}_{0s} - f_{s} \right), \\
 \partial _{x}E_{x} &=& \frac{e}{\varepsilon _{0}}\sum_{s}Z_{s}n_{s}, \\
 m_{s}\partial _{t}v_{ys} &=& eZ_{s}E_{y}, \\  
 \left( \partial _{t}\pm c\partial _{x}\right) E^{\pm} &=& 
   -\frac{e}{\varepsilon _{0}}\sum_{s}Z_{s}n_{s}v_{ys}, \\
 E^{\pm } &\equiv& E_{y}\pm cB_{z}.   
 \end{eqnarray}
$m_s$, $Z_se$, and $n_s\equiv\int dv_x\ f_s$ are the mass, charge, and number density of species $s$; $s=e$ for electrons, $Z_e=-1$, and $e>0$ is the positron charge. A number-conserving Krook relaxation operator is included, with
relaxation rate $\nu_{Ks}(x)$ and equilibrium distribution
$\hat{f}_{0s}$ set to the initial Maxwellian ($\int dv_x\,\hat{f}_{0s}=1$). We use a large $\nu_{Ks}=0.2\omega_p$ ($\omega_p^2\equiv n_0e^2/(\epsilon_0 m_e)$) at the edges of the finite length density profile to absorb plasma waves
generated by SRBS and to reduce fluctuations generated as the electron density adjusts to a sheath-like pattern. A nonzero central value of $\nu_{Ks}$ can mimic transverse sideloss from a laser speckle, although $\nu_{Ks}=0$ outside the damping regions in this paper.

The Vlasov equation for the distribution function $f_s$ is solved on a fixed phase-space grid via operator splitting \cite{cheng-vlasov-jcp-1976}. The spatial and velocity advections are performed via a semi-Lagrangian method (shift along characteristics) with cubic spline interpolation. The $x$ shift is periodic with $f_s(x_L)=f_s(x_R)$, although the dynamics in the central region are effectively finite in $x$ due to the damping regions and sheaths at the edges (subscripts $L$ and $R$ denote quantities at the left and right edges, respectively; see Fig.~\ref{fig:model_geom}). For the $v_x$ shift, $f_s$ is assumed zero for $|v_x|>v_\mathrm{max}$, which means particles are lost when accelerated beyond $v_\mathrm{max}$. After the $v_x$ shift, at each $x$ gridpoint we add to $f_s$ an initial Maxwellian with a density chosen to cancel the change in number caused by the $v_x$ shift. The transverse light-wave variables $E^+$ and $E^-$ advect to the right and left, respectively. We advance $E^{\pm }$ without dispersion (in vacuum) by shifts of one $x$ gridpoint, which locks the $x$ and $t$ spacings by $dx=c~dt$. The extreme edges of the simulation box, beyond the periodicity points $(x_L,x_R)$ of $f_s$, are ``moat'' regions where $f_s\equiv0$ for all time, but $E^\pm$ propagate until leaving the system. We specify $E^\pm(t)$ at the boundary from which each field is advected into the box, and can thereby inject a (pump, seed) light wave via $(E^+,E^-)$.

Due to the low numerical noise in our Vlasov code, SRBS does not develop if only the pump light wave is incident and the damping regions are used (although SRBS can occur with no damping regions). We therefore include a low-amplitude SRBS seed with a frequency that satisfies linear phase-matching based on the kinetic EPW dispersion relation. The physics is similar in runs where the seed was turned off after kinetically-enhanced SRBS develops.

\section{Intensity Scaling of SRBS  for Trident Conditions}
\label{sec:tridgen}

Simulations with conditions similar to the Trident single hot spot experiments \cite{montgomery-trident-pop-2002} reveal a sudden onset of kinetically-enhanced SRBS as the pump strength increases. A rapid increase in reflectivity was also observed in the experiments. We use a pump vacuum wavelength $\lambda_0=527$ nm, background electron density $n_0=0.025n_c$ ($\omega_0/\omega_p=6.32$) where $n_c$ is the pump critical density ($n_0/n_c=(\omega_p/\omega_0)^2$), and an electron temperature $T_e=0.5$ keV. The central flattop (between the Krook damping regions) has length 75.4 $\mu$m. We inject an SRBS seed with $\lambda_{1s}=653.4$ nm ($\omega_{1s}/\omega_p=5.10$) and intensity $I_{1R}=10^{-5}I_{0L}$ at the right edge, where $I_{0L}$ is the pump intensity at the left edge (in this paper, a constant $I_0$ independent of $x$ and $t$ is understood to be $I_{0L}$). The pump and seed light waves beat to drive an EPW with $k_2\lambda_D=0.352$. In all runs we set $dx=\lambda_2/20$ and $v_x$ grid spacing $dv_x<v_{tr,s}/4$. $v_{tr}\equiv2\omega_B/k_2$ is the trapping island half-width in the EPW electric field, $\omega_B\equiv\omega_p(n_2/n_0)^{1/2}$ is the trapped electron bounce frequency, and $n_2$ is the EPW density perturbation peak amplitude. $v_{tr,s}$ is $v_{tr}$ for the EPW produced by beating of the pump and unamplified seed light waves; in other words, $dv_x$ resolves trapping even if there were no seed amplification. The $v_x$ grid extends to $v_\mathrm{max}=(6.4,8.6,9.6)v_{Te}$ for $I_{0L}(<,=,>)1$ \pw.

\begin{figure}
  \includegraphics{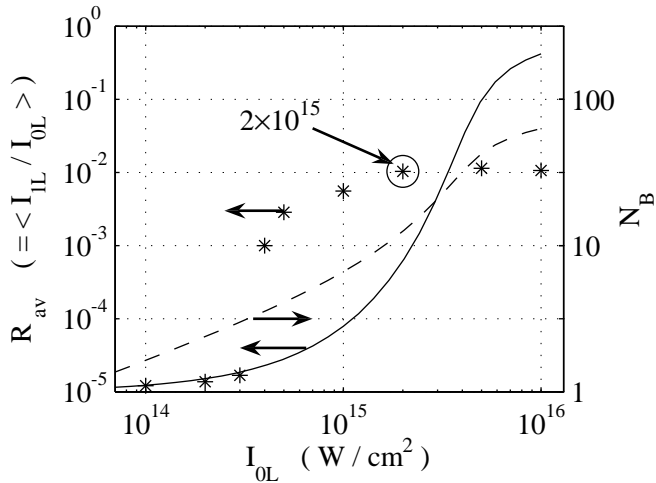}
  \caption{\label{fig:trid_refavg} Time-averaged reflectivity \rav\ and bounce number $N_B$ for Trident parameters described in text. Asterisks are simulation \rav. The solid curve is the coupled-mode steady-state \rav\ from Eq.~(\ref{eq:TangR}), and the dashed curve is $N_B$ from Eq.~(\ref{eq:NBpd}); both include pump depletion.}
\end{figure}

Fig.~\ref{fig:trid_refavg} shows $\rav$ versus pump strength, where $\rav$ is the reflectivity $R\equiv\left<I_{1L}\right>/I_{0L}$ time-averaged starting from 1 ps (to allow the lasers to turn on and transit the domain) to the end of the run (at 10 ps). All runs were below the absolute instability threshold $I_{0a}=29.8$ \pw\ for undamped light waves, which corresponds to the threshold undamped SRBS growth rate $\gamma_{0a}=(1/2)|v_{g1}/v_{g2}|^{1/2}\nu_2$, with $\nu_2$ the EPW Landau damping rate. Coupled-mode theory thus predicts SRBS approaches a temporal steady state where the seed light wave is convectively amplified across the box. The steady state can be solved analytically for a three-wave model that includes pump depletion but neglects light-wave damping, and neglects EPW advection compared to damping (the strong damping limit) \cite{tang-sbs-jap-1966}. The strong damping limit is valid for $I_{0L}\ll I_{0a}$. The scattered light wave in this limit satisfies
\begin{equation}
  {d\hat{I_1} \over dx} = -{G\over L} \hat{I}_0\hat{I}_1
\end{equation}
where $\hat{I}_0=I_0(x)/I_{0L}$, $\hat{I_1}=I_1(x)/I_{1R}$, and $I_0-(\omega_0/\omega_1)I_1=$ constant. $L$ is the length of the gain region, and the intensity gain exponent $G$ in practical units is 
 \begin{equation}
 G \equiv {\lambda_{1p}\over\lambda_2}(1-n_0/n_c)^{-1/2}\mathrm{Im}\left[{\chi\over 1+\chi}\right] {I_{0L,\mathrm{pw}}\lambda_{0,\mu}^2 \over 871}{L\over\lambda_2}.
 \end{equation}
$\lambda_{1p}$ is the scattered light wavelength in the plasma, $\lambda_{0,\mu}$ is $\lambda_0$ in microns, and $I_{0L,\mathrm{pw}}$ is $I_{0L}$ in \pw. $\chi(k_2,\omega_2)=-(k_2\lambda_D)^{-2}Z'[\omega_2/(k_2v_{Te}\sqrt2)]$ is the electron susceptibility, where $Z$ is the plasma dispersion function.  $k_2$ and $\omega_2$ are given by the beating of the chosen pump and scattered light waves. The resulting reflectivity is
 \begin{equation}
 \label{eq:TangR}
 \tilde{R}(1-\tilde{R}+\tilde{s}) = \tilde{s}\exp \left[ G(1-\tilde{R}) \right]
 \end{equation}  
with $\tilde{R}=(\omega_0/\omega_1)R$ and $\tilde{s}=(\omega_0/\omega_1)I_{1R}/I_{0L}$.

For weak pumps the simulation gains are small and in accord with the coupled-mode calculation, but they suddenly rise well above it for $I_0=0.4$ \pw\ (kinetic inflation or enhancement). \rav\ barely increases with pump strength and saturates around 1\% for $I_0\gtrsim 1$ \pw. This is well below the coupled-mode level for the largest pump strengths and indicaties nonlinearities besides pump depletion saturate SRBS (kinetic deflation).

We do not expect trapping to be important in SRBS unless a resonant electron undergoes at least one bounce cycle before crossing the gain region. This happens for all the runs in Fig.~\ref{fig:trid_refavg}, using the convective steady-state fields. The above criterion treats longitudinal endloss as the only de-trapping mechanism, since we have not included transverse sideloss (e.g., via the Krook operator) in our runs. Morales and O'Neil calculate the time-dependent damping rate and frequency of an undriven EPW as an initial value problem \cite{oneil-damping-pof-1965, morales-freqshift-prl-1972}. That is, the damping and frequency vary with the number of bounce orbits the trapped particles have experienced, defined below as $N_B$. They find that after the resonant electrons complete about one bounce cycle, Landau damping is substantially reduced, and the frequency shift is roughly its late-time asymptotic value. These results easily translate to a boundary value problem, appropriate to our finite geometry. However, driven EPW's in SRBS may differ from Morales and O'Neil's free-wave calculation: see the discussion at the end of this section.

In steady state, resonant electrons emerge from a Maxwellian distribution at the left damping region ($x=0$) and undergo $N_B(x)$ bounce orbits (dependent on the field amplitude) before reaching the position $x$. Across a domain of interest $[0,x]$,
 \begin{equation}
 N_B(x) \equiv (2\pi)^{-1}\int_0^x dx'\ k_B(x')
 \end{equation}
where $k_B\equiv\omega_B/v_{p2}$ is the bounce wavenumber in the local field $E_x(x)$. The total amplitude (drive plus response) of the EPW driven by two beating light waves as given by the linearized Vlasov equation is
\begin{equation}
{n_2 \over n_0} = {1\over2}(k_2\lambda_D)^2 {1 \over 1+\chi} {v_{\mathrm{os0}}v_{\mathrm{os1}}^* \over v_{Te}^2}.
\end{equation}
$v_{Te}\equiv(T_e/m_e)^{1/2}$ and $v_{osi}\equiv eE_i/(m_e\omega_i)$ is the complex electron oscillation velocity in light wave $i$. A Fourier amplitude $f$ is related to its physical field $f_p$ by $f_p=(1/2)f\exp i(kx-\omega t)+\mathrm{cc}$. In the convective steady state described above, $n_2 = n_{2s}(\hat{I}_0\hat{I}_1)^{1/2}$ where $n_{2s}$ is $n_2$ for $I_0=I_{0L}$ and $I_1=I_{1R}$. The resulting $N_B$ is
\begin{equation}
  \label{eq:NBpd}
  N_B(x) = N_{Bs}(x){1\over x}\int_0^x dx'\ (\hat{I}_0\hat{I}_1)^{1/4}
\end{equation}
where $N_{Bs}(x) \equiv [\omega_p x/(2\pi v_{p2})](n_{2s}/n_0)^{1/2}$ is $N_B(x)$ if there were no seed amplification. Neglecting pump depletion for illustration, $v_{\mathrm{os1}} = v_{\mathrm{os1},R}\exp[(G/2)(1-x/L)]$, and $N_{BR}$, or the total $N_B$ across the domain of length $L$, is
 \begin{equation}
 \label{eq:NBnopd}
  N_{BR} = N_{Bs}(L){e^{G/4}-1 \over G/4}.
 \end{equation}
Note that $N_{Bs}\propto I_{1R}^{1/4}$ depends very weakly on the seed intensity.

$N_{BR}$ in the convective steady state is shown in Fig.~\ref{fig:trid_refavg} as a function of pump strength. Substantial bouncing occurs even for the weakest pump ($N_{BR}$=1.63 for $I_0$=0.1 \pw), suggesting that the convective profiles may be altered by a damping rate and frequency that vary appreciably across the plasma with $N_B(x)$. However, kinetic enhancement only occurs for $N_{BR} \gtrsim 3.5$, with $N_{BR}$=3.52 for the smallest enhanced run ($I_0$=0.4 \pw). This runs counter to the simple-minded expectation, based on the Morales-O'Neil free-wave theory, that SRBS inflation should occur for $N_{BR}\approx 1$ (or even less, since the damping rate in Fig.\ 1 of \cite{morales-freqshift-prl-1972} first becomes zero for $N_B\approx0.6$). This may indicate a difference between the damping rates for free and driven waves. It could also be that inflation requires the damping to be small over a sufficiently large region, and not just near the right edge.  These questions merit further research.

\section{Trident case $I_0=2$ PW/cm$^2$: Kinetic enhancement and Electron Acoustic Scatter }
\label{sec:trid2e15}

\subsection{Reflected Light}
\begin{figure}
 \includegraphics{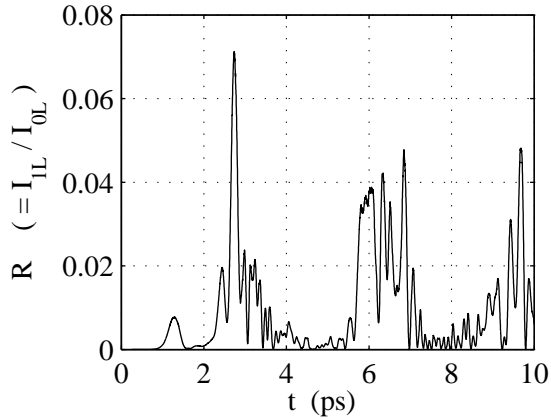}
 \caption{\label{fig:trid_reft} Instantaneous reflectivity $R$ for $I_0=2$ \pw\ Trident run.}
 \end{figure}

We now study in detail the case $I_0=2$ \pw\ to expose the physics of kinetic enhancement and EAS. Figure \ref{fig:trid_reft} displays the instantaneous reflectivity, with time average $\rav=1.03\%$ that greatly exceeds the coupled-mode level from Eq.~(\ref{eq:TangR}) of $7.75\times10^{-4}$. Moreover, $R(t)$ occurs in temporal sub-picosecond bursts instead of becoming steady, as has also been reported in PIC simulations (e.g., Refs.~\cite{vu-kininf-prl-2001, langdon-rescat-prl-2002}).

\begin{figure}
  \includegraphics{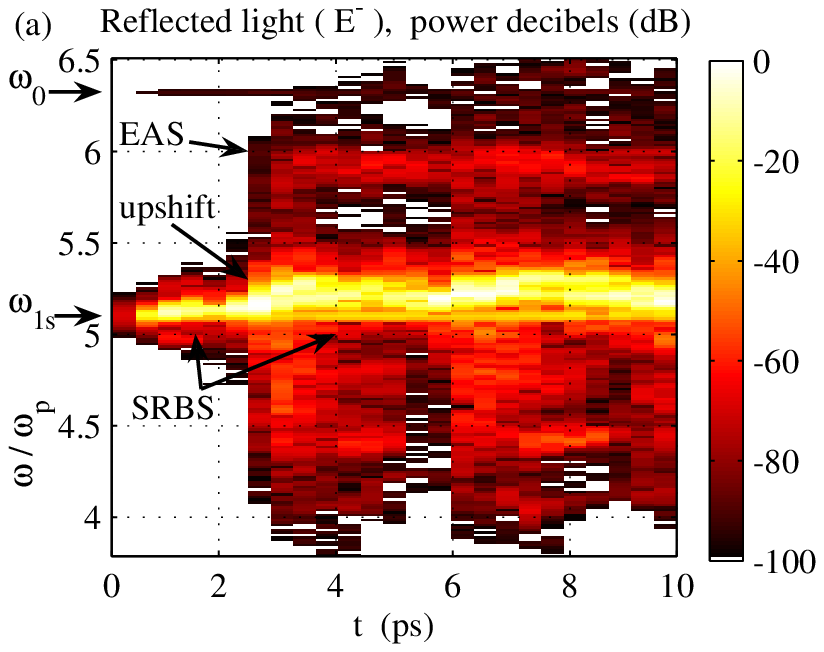} \\
  \includegraphics{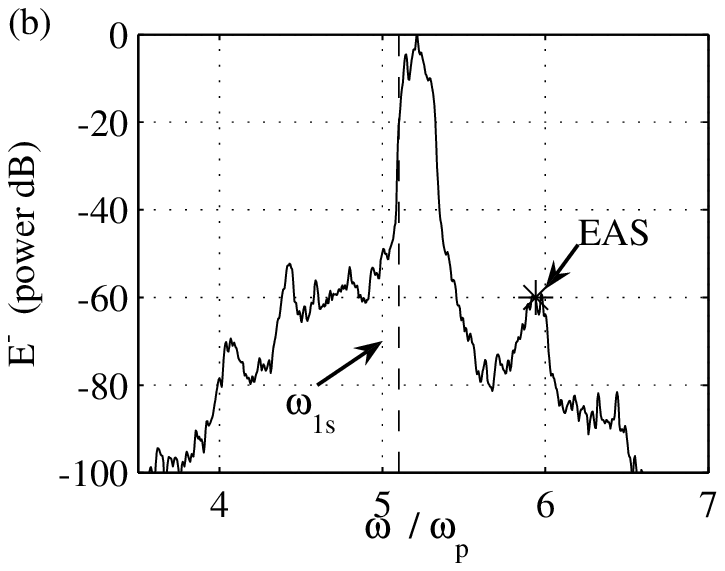}
  \caption{ \label{fig:trid_emwall} (Color.) (a) Time-resolved and (b) time-integrated spectrum of reflected light $E^-$ for $I_0=2$ \pw\ Trident run. ($\omega_0$, $\omega_{1s}$) are the imposed (pump, seed) frequency. The $\omega$ spectrum of a field $f$ in ``power dB'' is $20\log_{10}|f(\omega)|$. The EAS asterisk in (b) is at $\omega=5.94\omega_p$.}
\end{figure}

The time-resolved spectrum of reflected light is displayed in Fig.~\ref{fig:trid_emwall}(a). SRBS is the dominant signal. Early in time it occurs at the seed frequency $\omega_{1s}$ but upshifts once SRBS becomes strongly enhanced around 3 ps. This results from the trapping-induced EPW frequency downshift \cite{morales-freqshift-prl-1972}. Scattered light between the SRBS and pump frequencies is observed shortly after the upshift. The time-integrated power spectrum, contained in Fig.~\ref{fig:trid_emwall}(b), reveals EAS is much stronger than the ``noise level'' in neighboring frequencies (the noise is partly due to SRBS sidelobes resulting from windowing). The EAS reflectivity is $R_\mathrm{eas}$=2.06$\times10^{-8}$, from the reflected light in the band $\omega = (5.8-6.05)\omega_p$ during the time EAS is observed ($t>2$ ps). The broad spectrum of reflected light for $\omega<\omega_{1s}$, larger than the signal above the SRBS peak, is scattering off electrostatic fluctuations (possibly from BAMs) with $\omega>\omega_2$ visible in Figs.\ \ref{fig:trid_exwk_all}(a) and \ref{fig:trid_exwk_roots}.

All power spectra in this paper were made using Welch's method of averaged overlapping modified periodograms, with Kaiser window functions \cite{numrec, proakis-signal-2002}. Aggressive windowing was needed to reveal weak, time-dependent signals, such as EAS in the presence of much larger SRBS. The low noise inherent in Vlasov codes allows such weak processes to be captured.

\subsection{Electrostatic Activity}
\label{sec:trides}

\begin{figure}
 \includegraphics{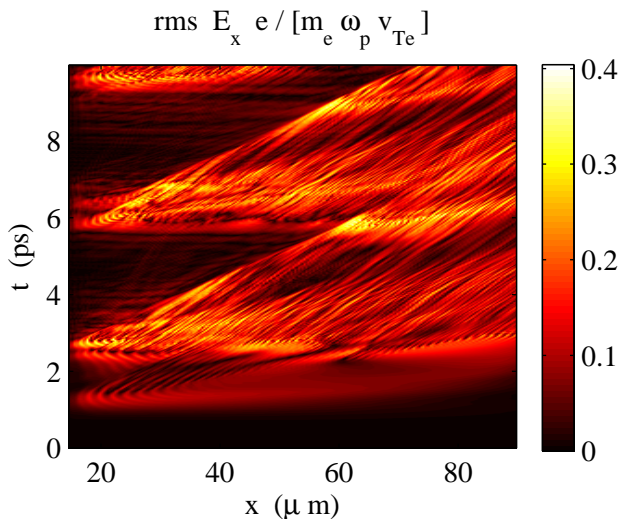}
 \caption{ \label{fig:trid_extx} (Color.) Envelope (rms) of electrostatic field $E_x(x,t)$ for $I_0=2$ \pw\ Trident run over the $x$ domain between the damping regions.}
 \end{figure}

We now turn to the electrostatic activity. Fig.~\ref{fig:trid_extx} presents the envelope of the longitudinal electric field, found by rms-averaging $E_x(x,t)$ over one SRBS wavelength and period. Once SRBS enters the enhanced regime, the plasma waves propagate away from the laser entrance along fairly well-defined ``rays'' with group velocities $\approx v_{Te}$. Three broad ``bursts,'' which are strong near the laser entrance around 3, 6, and 10 ps, correspond to reflectivity bursts in Fig.~\ref{fig:trid_reft}. The initial break-up of the envelope into rays, near the laser entrance shortly after 1 ps, may be due to the trapped particle instability \cite{kruer-tpi-prl-1969, brunner-valeo-prl-2004}, and coincides with the first, isolated spike in reflectivity at 1.3 ps. In particular, the envelope is modulated at $k\lambda_D\approx0.05$, which is close to the bounce wavenumber $k_B\lambda_D=0.066$ calculated for the observed amplitude $E_xe/(m_e\omega_pv_{Te})\approx0.15$. Also, the electrostatic spectrum over the space-time region of the break-up (not shown) possesses sidebands at $(k_2,\omega_2)\pm(k_B,\omega_B)$. However, we do not see clear signs of the trapped particle instability for later times.

\begin{figure*}    
 \includegraphics{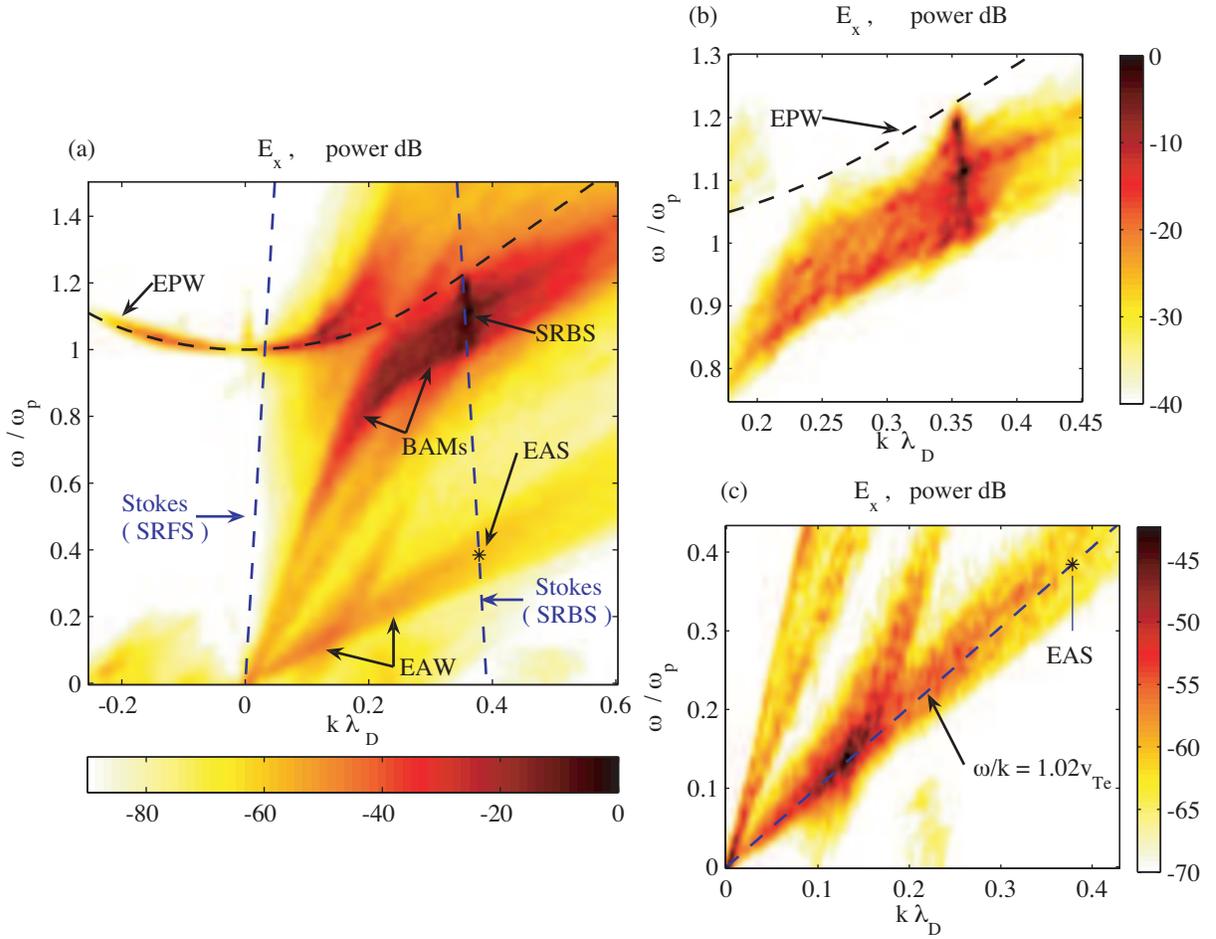}
 \caption{ \label{fig:trid_exwk_all} (Color.) Power spectrum of $E_x$ over the $(x,t)$ domain of Fig.~\ref{fig:trid_extx} for $I_0=2$ \pw\ Trident run. The ``EPW'' curve is the EPW for the initial Maxwellian, and the ``Stokes'' curve is defined in the text. The ``EAS'' point at $0.385\omega_p$ is phase-matched with the EAS asterisk in Fig.~\ref{fig:trid_emwall}(b). Panels (b) and (c) are zooms of the SRBS and EAW regions of panel (a).}
 \end{figure*}

The electrostatic $(k,\omega)$ spectrum, taken over all time and between the two Krook damping regions, is shown in Fig.~\ref{fig:trid_exwk_all}. The ``Stokes'' curve is the locus of daughter electrostatic modes $(k_2,\omega_2)$ phase-matched for electromagnetic decay of the pump, namely, $(k_2,\omega_2)=(k_0,\omega_0)-(k_1,\omega_1)$, with $\omega_1\in[0,\omega_0]$ and $ck_i=\pm(\omega_i^2-\omega_p^2)^{1/2}$ for $i=0,1$. $k_1>0$ and $k_1<0$ give, respectively, the smaller $k_2$ (forward SRS) and larger $k_2$ (backward SRS) legs of the curve. SRBS occurs along the Stokes curve, mostly on a frequency-downshifted streak relative to the matched EPW. While there is some activity on the EPW curve for $k<0$, the curve splits into upper and lower branches near $k\lambda_D=0.2$. The extended activity lower in frequency than the EPW curve, and which becomes acoustic ($\omega\propto k$) with a larger slope for small $k$, is analogous to the beam acoustic modes (BAMs) found by L.~Yin \textit{et al.} \cite{yin-sbams-pre-2006}.

A low-amplitude acoustic mode ($\omega\propto k$) with phase velocity $\approx v_{Te}$, which we call the electron acoustic wave (EAW), is also present. It is mostly energized for $k\lambda_D\approx0.15$, and remains acoustic but weakens toward higher $k$. The most strongly-excited EAW has a frequency $\approx0.15\omega_p$ well above the ion plasma and ion acoustic wave frequencies, so EAW interaction with ions can be neglected. Spectra over successive time windows reveal the EAW is first excited at, and is always strong for, lower $k$. The EAW and EAS both develop after kinetic enhancement of SRBS begins, which suggests they result from SRBS-induced modifications of the electron distribution. The asterisk at $(k\lambda_D,\omega/\omega_p)=(0.379,0.385)$ with phase velocity 1.02$v_{Te}$ is phase-matched to the EAS asterisk at $\omega=5.94\omega_p$ in Fig.~\ref{fig:trid_emwall}(b). However, the EAW curve is not generated by EAS, since most of the EAW energy is far from the phase-matched point. Also, the amplitude is not increased near the phase-matched point, implying decay of the pump laser does not significantly change the EAW level. The scattering from the EAW appears to be off independently-generated fluctuations and can be viewed as electron acoustic Thomson scatter (EATS) in analogy with Thomson scattering from conventional plasma waves.

We propose that a three-wave interaction involving different parts of the extended BAM feature generates the low-$k$ EAWs, and dub this process beam acoustic decay (BAD). The rest of the EAW curve is weakly excited by EAW harmonic coupling (harmonics of an acoustic mode lie on the same acoustic curve). BAD is likely a two-pump process (instead of a pure parametric decay), where the BAM curve is energized separately from BAD, and BAMs with different $(k,\omega)$ then beat to produce EAWs. This is supported by the fact that the low-$k$ BAM tail does not have a noticeable peak corresponding to the daughter BAM in BAD. Moreover, one can estimate the ratio of daughter-wave amplitudes expected from coupled-mode theory. For weakly damped BAMs and a strongly-damped EAW (appropriate for the linear modes found in Sec.\ \ref{subsec:modes}), and neglecting pump depletion, the ratio of daughter BAM to daughter EAW actions is $(\nu_e/\gamma_b)^2$ where $\nu_e$ is the EAW damping rate and $\gamma_b$ is the BAD growth rate. For undamped daughters the actions are equal (the traditional Manley-Rowe relation). The $E_x$ power ratio from Fig.~\ref{fig:trid_exwk_all} is very large: $|E_{x,\mathrm{bam}}/E_{x,\mathrm{eaw}}|^2 \approx 10^4$. Still, without formulae for $\gamma_b$ and how electric fields relate to actions, it is difficult to infer that little of the BAM is produced by BAD. We discuss the generation of BAMs below. 

The EAW and resulting EAS occur for moderate pump strengths. Neither appear when SRBS is not kinetically enhanced. Both increase with $I_0$ above the enhancement threshold, until no distinct EAS peak is visible for $I_0\geq5$ \pw. The upshift in $\omega_1$ due to the nonlinear downshift of the SRBS EPW grows significnatly with pump strength. As the pump increases, there is also more reflected light higher in freqnuency than SRBS; however, this light is spectrally broad and has no peaks (such as EAS). A distinct EAW is seen for the $I_0=5$ \pw\ run but not for $I_0=10$ \pw. We discuss the relation of our results to experimental reports of SEAS in the conlcusion.

\subsection{Distribution function and linear modes}
\label{subsec:modes}
\begin{figure}
\includegraphics{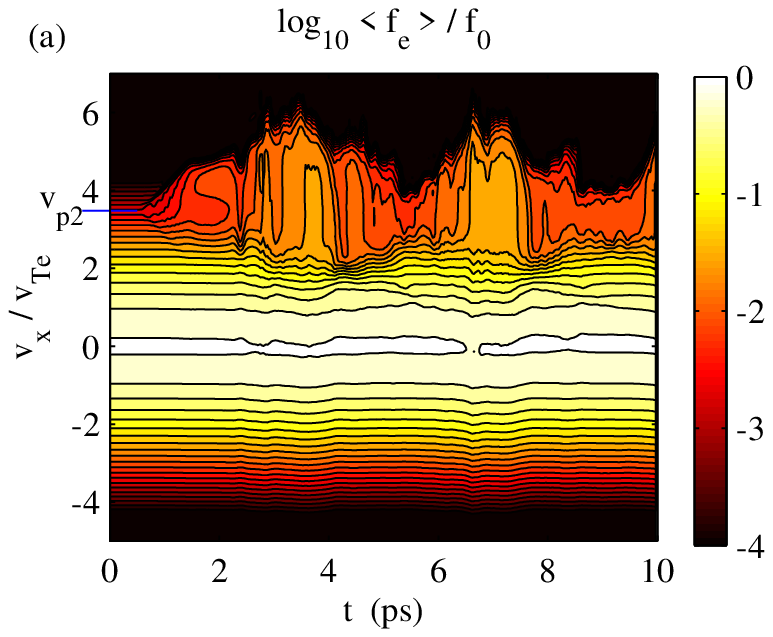} \\
\includegraphics{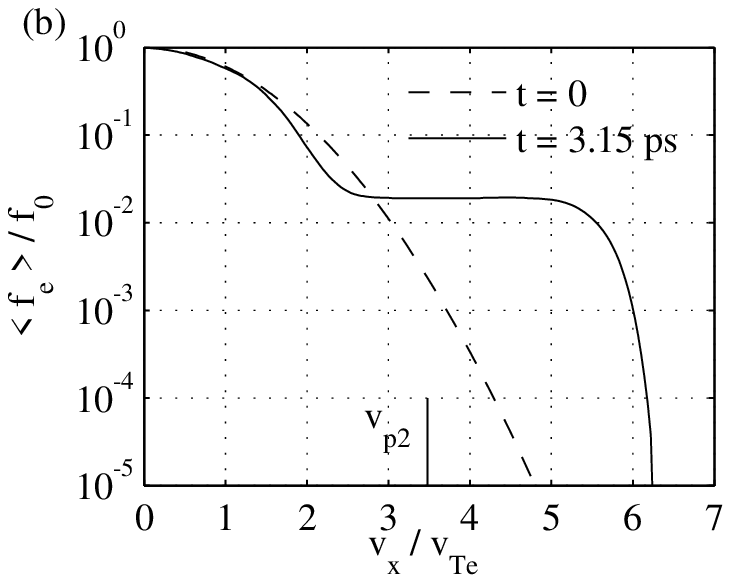}
\caption{\label{fig:trid_ftplog} (Color online.) $f_e$ space- and time-averaged over $\omega_p\Delta t=6.3$ and $x=$ 50.6-53.8 $\mu$m for $I_0=2$ \pw\ Trident run. Panels (a) and (b) are, respectively, for all times and $t=$ 0, 3.15 ps. $v_{p2}=3.48v_{Te}$ is the phase velocity of the EPW driven by the beating of the pump and seed light waves. $f_0\equiv n_0/(v_{Te}\sqrt{2\pi})$.}
\end{figure}

We now examine the electron distribution $f_e$ and the linear electrostatic modes it supports. Phase-space vortices develop in the electric field of the SRBS plasmon and flatten the space-averaged $f_e$ near $v_{p2}$. Although the vortices are regular for early times, once SRBS becomes enhanced they smear into each other.  This indicates a non-monochromatic spectrum and has been seen to result, e.g., from the trapped particle instability in Refs.~\cite{kruer-tpi-prl-1969, brunner-valeo-prl-2004}. Figure~\ref{fig:trid_ftplog}(a) displays $f_e$ space- and time-averaged over periods of 6.3$/\omega_p$ and from 50.6-53.8 $\mu$m (near the domain center) for the $I_0=2$ \pw\ Trident run. The plateau width is somewhat correlated to the local wave amplitude, but $f_e$ remains flattened even when the field is weak. This suggests nonlocality may be important in kinetically-enhanced SRBS. 

The averaged $f_e$, taken from Fig.\ \ref{fig:trid_ftplog}(a) at 3.15 ps, is plotted in Fig.~\ref{fig:trid_ftplog}(b) and is fairly representative of $f_e$ during enhanced SRBS. The observed flattened plateau half-width (for the time and space region in Fig.~\ref{fig:trid_ftplog}(b)) of 1.5$v_{Te}$ compares well with $v_{tr}=1.47v_{Te}$ from the observed amplitude $E_xe/(m_e\omega_pv_{Te})=0.192$. Note that $f_e(v_x)$ in the trapping region is nearly flat (it actually increases by 1.6\% over its minimum for $v_x\gtrsim v_{p2}$). This is typical for fully-enhanced SRBS, although in the first Raman burst there is a clear period when $df_e/dv_x>0$ for $v_x\gtrsim v_{p2}$. Such an upturn appears as the pinched bulge in Fig.~\ref{fig:trid_ftplog} near $(t,v_x)$=(2 ps, 4$v_{Te}$). Some adiabatic calculations of trapping in EPWs show that $f_e$ should be symmetric about the phase velocity when the wave amplitude decreases and electrons are detrapped \cite{benisti-adiabatic-pop-sub2006}. 

\begin{figure}
\includegraphics{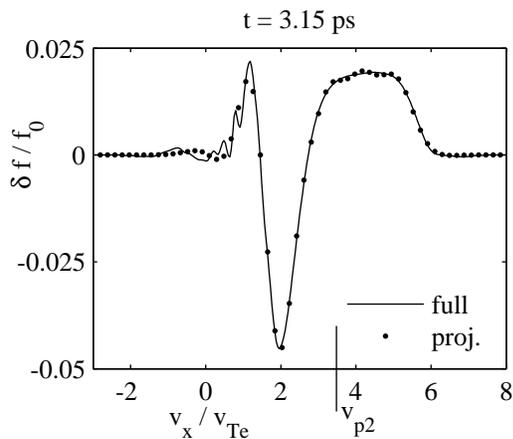}
\caption{\label{fig:trid_df} $\delta f = f_e-f_M$ (solid) and its Gauss-Hermite projection (dotted) for $\left<f_e\right>$ from Fig.\ \ref{fig:trid_ftplog} at $t=3.15$ ps. Projection parameters given in text.}
\end{figure}

Figure~\ref{fig:trid_df} shows $\delta f=f_e-f_M$ at $t=3.15$ ps, where $f_M$ is a Maxwellian fit to $f_e$ for $v_x\in[-4,0]v_{Te}$ (close to the initial Maxwellian). Note $\delta f$ consists of a broad, beam-like structure near $v_{p2}$ due to trapping, as well as at two lower velocities. Modeling $f_e$ as a bulk-and-beam distribution, such as a bi-Maxwellian, may not be correct, especially for modes like EAWs with low phase velocities near the secondary structures.

To gain insight into the electrostatic dynamics, we study the linear modes of the numerically-obtained $f_e$. Although a large-amplitude electric field is present and may demand a nonlinear calculation, linear theory yields good agreement with the observed spectrum. We project $\delta f$ onto a Gauss-Hermite basis, as described in the Appendix. This gives an analytic longitudinal susceptibility $\chi$, valid in the complex plane, as the sum of $\chi$ for $f_M$ and the projected $\delta f$. Figure \ref{fig:trid_df} displays the full $\delta f$ and its projection up to the $N=12$ basis function and with the scaled, shifted velocity $u=(v/v_{Te}-3)/0.7$. The projection only misses fine-scale features which would appear in higher-order basis functions. These features should weakly distort the real frequency, which stems from collective oscillations in $f_e$, but may affect the damping rate more strongly, which depends on the slope of $f_e$ at the phase velocity.

\begin{figure}
\includegraphics{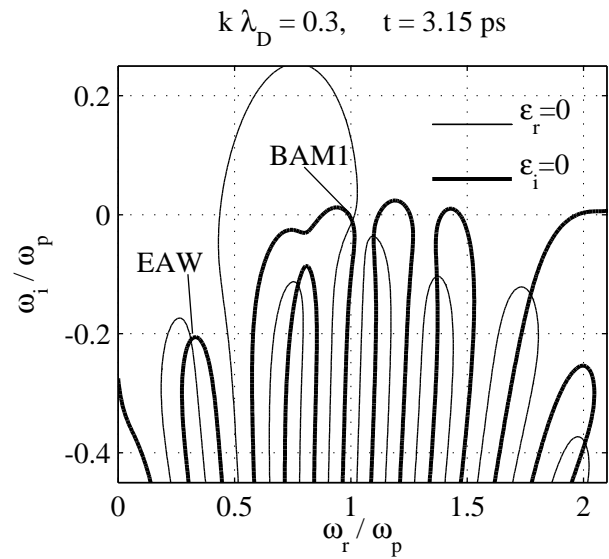}
\caption{\label{fig:trid_chi} Zero contours of the electrostatic dispersion relation $\epsilon=0$ for $k\lambda_D=0.3$, computed from $f_e$ as projected in Fig.~\ref{fig:trid_df}.}
\end{figure}

The electrostatic linear modes are obtained by solving the dispersion relation $\epsilon\equiv1+\chi=0$ for complex $\omega$ given real $k$. We do not include a ponderomotive term due to parametric coupling with light waves, which would allow for SRBS. Fig.~\ref{fig:trid_chi} presents the zero contours of $\epsilon_r$ and $\epsilon_i$, which intersect at the linear modes $\epsilon=0$, for $f_e$ from Fig.\ \ref{fig:trid_df}. This plot over the same $(k,\omega)$ region only using $f_M$ has two zero contours, which intersect at the linear EPW; there is an infinity of strongly-damped modes for the Maxwellian, which fall below the chosen plot range. As $N$ is increased the contours change slightly, and a single pair of $\epsilon_r=0$ and $\epsilon_i=0$ contours can split into two pairs. However, the plot remains qualitatively the same. This is further discussed in the Appendix.

\begin{figure}
\includegraphics{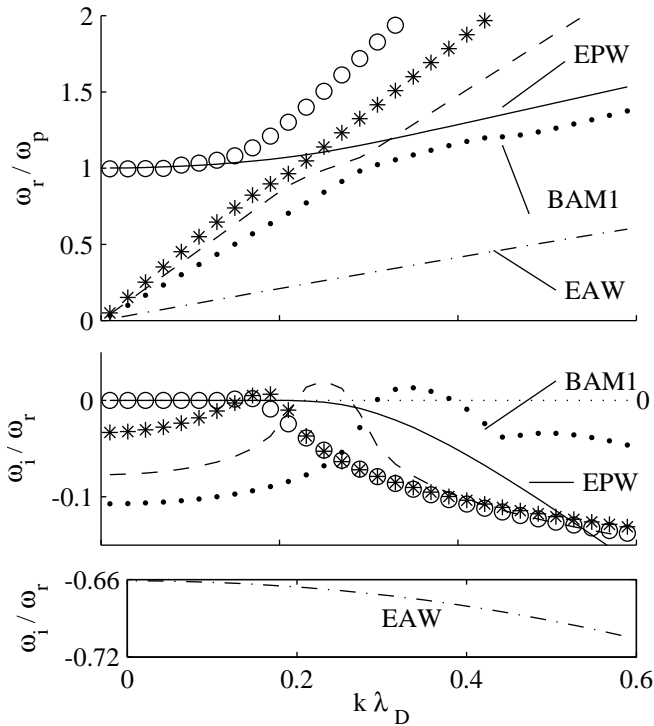}
\caption{\label{fig:trid_froots} Electrostatic linear modes from the projection in Fig.~\ref{fig:trid_df}. ``EPW'' indicates the EPW for the initial Maxwellian.}
\end{figure}

\begin{figure}
\includegraphics{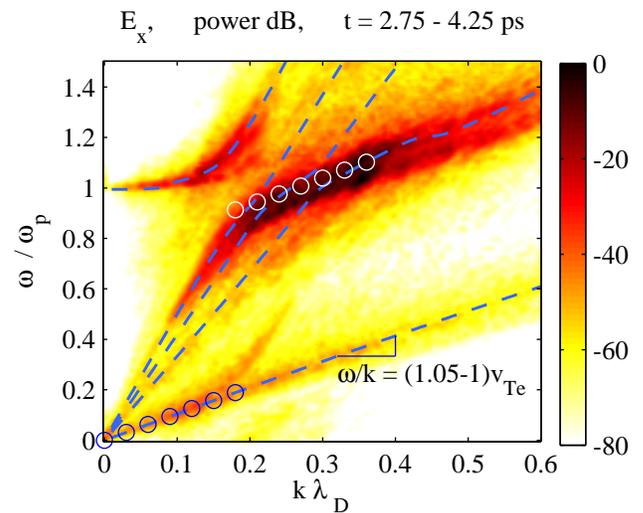}
\caption{\label{fig:trid_exwk_roots} (Color.) Roots from Fig.~\ref{fig:trid_froots} overlaid on $E_x(k,\omega)$ for $t=2.75-4.25$ ps and the $x$ domain of Fig.~\ref{fig:trid_exwk_all}. The white and blue circles are possible phase-matched daughter BAMs and EAWs, respectively, for beam acoustic decay of a parent BAM1 root with $(k\lambda_D,\omega/\omega_p)=(0.36,1.1)$.}
\end{figure}

Figure~\ref{fig:trid_froots} graphs five modes, calculated for $f_e$ from Fig.\ \ref{fig:trid_df} ($t=3.15$ ps), that resemble the observed spectrum. In Fig.~\ref{fig:trid_exwk_roots}, $\omega_r$ for these modes is superimposed on $E_x(k,\omega)$ computed for $t=2.75-4.25$ ps; the agreement is quite good. The linear modes show the splitting of the EPW curve into an upper branch and a set of BAMs, as well as the appearance of an EAW. SRBS occurs along the mode labeled BAM1, which is frequency-downshifted and less damped than the Maxwellian EPW. Thus, much of the nonlinear EPW physics of Morales and O'Neil is recovered by linear theory with the modified $f_e$, as also seen in \cite{williams-nonlinsbs-pop-2004}.

The upper branch and BAMs are linearly unstable for some $k$, \textit{in the absence of coupling to light waves.} Since the slope of the modified $f_e$ is small or slightly positive for some $v_x$, it is reasonable to find bump-on-tail or beam-plasma instabilities. These beam-driven, growing electrostatic modes have been studied previously \cite{oneil-beamroots-pof-1968, gary-elinst-jgr-1985}, and may account for the higher BAM fluctuations compared to the level of $k<0$ conventional EPWs in Fig.~\ref{fig:trid_exwk_all}. The linear instability of BAMs allows them to be excited separately from BAD, which based on the discussion at the end of Sec.\ \ref{sec:trides} seems to have little effect on the BAMs.

The linear modes also contain an EAW with $\omega/k$  decreasing from $1.05v_{Te}$ to $v_{Te}$ as $k$ increases. It is always heavily damped, unlike the undamped nonlinear EAW due to trapping discussed by other workers (e.g., Refs.\ \cite{schamel-holeq-pop-2000, rose-nonlinEPW-pop-2001}). An EAW also appears in an analogous mode calculation for a bi-Maxwellian $f_e$ presented in Ref.~\cite{salcedo-srs-nf-2003}, with a beam whose drift velocity is close to that of the EAW.  It is shown there that stimulated (quasi-mode) scattering off the EAW has a small, positive growth rate. Possible phase-matched BAD triplets are shown in Fig.~\ref{fig:trid_exwk_roots}. Note that the lowest-$k$ daughter BAMs (white circles) lie above the ``knee'' in the BAM feature, and match with daughter EAWs (blue circles) higher in $k$ than the strongest EAW activity. The break in the BAM streak may thus explain the cutoff in the EAW spectrum, although a calculation of the nonlinear coupling mechanism (e.g., coupling coefficients) would shed more light.

\subsection{Bispectral Analysis}

To further support the three-wave processes of BAD and EATS, we turn to bispectral analysis \cite{nikias-bispec-procieee-1987,mendel-hos-procieee-1991}. This has been used previously to study three-wave interactions in plasma physics, mostly in magnetized plasmas (Ref.\ \cite{kim-bispec-ieeetps-1979} is an early example). Although we do not pursue it here, bispectral techniques can give estimates of the coupling coefficients and mode growth rates \cite{ritz-3wi-pofb-1989,burin-turbulence-pop-2005,sen-feedback-pop-2000}.

The power spectrum, or the Fourier transform of the two-point correlation function, can be generalized to correlations among more signals at more times, or higher-order spectra. Given three real, zero-mean signals $x(t)$, $y(t)$, and $z(t)$, their correlation function $C_3$ and complex bispectrum $P_3$ are
\begin{equation}
  C_3\{x|y,z\}(\tau_1,\tau_2) \equiv \int dt\, x(t)y(t+\tau_1)z(t+\tau_2)
\end{equation}
and
\begin{equation}
  P_3(\omega_1,\omega_2) \equiv \int d\tau_1d\tau_2\, e^{i(\omega_1\tau_1+\omega_2\tau_2)}C_3(\tau_1,\tau_2).
\end{equation}
The notation $\{x|y,z\}$ indicates the fields being correlated, and Eq.~(\ref{eq:P3j}) shows why $x$ is treated differently. We compute the bispectrum by a generalization of Welch's method for power spectra \cite{proakis-signal-2002}. Namely, we divide time into $N$ windows, apply a window function to the signals, find their Fourier transforms, and in each window form
\begin{equation}
  \label{eq:P3j}
  P_{3j}(\omega_1,\omega_2) = X_j^*(\omega_1+\omega_2)Y_j(\omega_1)Z_j(\omega_2).
\end{equation}
$P_3 = \left\langle P_{3j} \right\rangle$ where $\left\langle\right\rangle$ denotes averaging over the $N$ windows. $P_3$ measures the signal amplitudes that frequency match (as three-wave interactions do). If the signals at $\omega_0\equiv\omega_1+\omega_2$, $\omega_1$, and $\omega_2$ have independent phases $\phi_i$ but constant amplitudes, then the $P_{3j}$'s have random phase factors and $|P_3|\sim N^{1/2}$. This corresponds to three independent signals that ``gratuitously'' frequency match, although they are not dynamically coupled. However, for a perfectly coherent process like a pure three-wave interaction, $\phi_0=\phi_1+\phi_2$, the $P_{3j}$'s are in phase, and $|P_3|\sim N$. The degree of coherence is quantified by the bicoherence $b_3$:
\begin{equation}
  b_3 \equiv {P_3 \over \left\langle \left| P_{3j} \right|^2 \right\rangle^{1/2}}.
\end{equation}
$|b_3|$ always lies between zero (no phase coherence) and one (perfect phase coherence), and its denominator scales like $N$.

\begin{figure}
\includegraphics{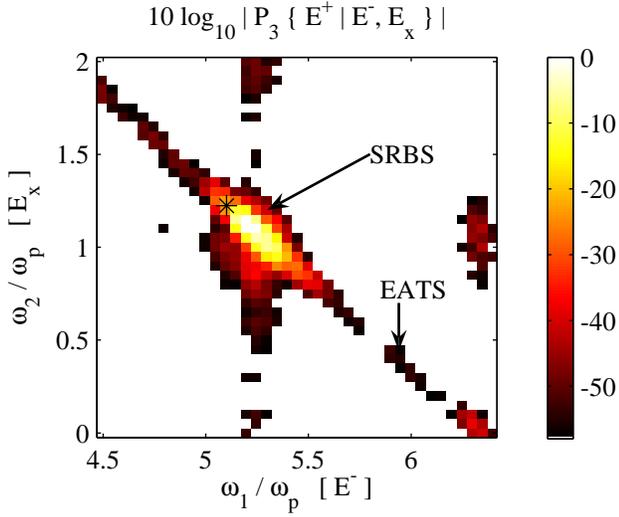}
\caption{\label{fig:bispec_srs_2d} (Color online.) Bispectrum $|P_3\{E^+|E^-,E_x\}|$ for $t=5-10$ ps at $x=52.4\ \mu$m for $I_0=2$ \pw\ Trident run. The asterisk marks the seeded linear SRBS point. $\omega_1+\omega_2=\omega_0$ along the dominant diagonal streak.}
\end{figure}

\begin{figure}
\includegraphics{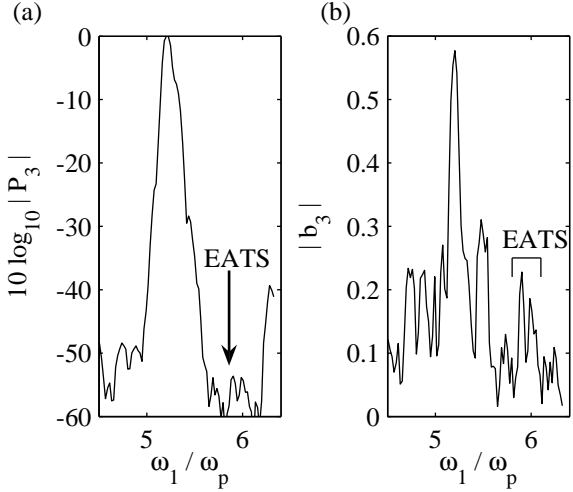}
\caption{\label{fig:bispec_srs_1d} (a) bispectrum $|P_3|$ and (b) bicoherence $|b_3|$ from Fig.~\ref{fig:bispec_srs_2d} along the line $\omega_1+\omega_2=6.32\omega_p$ for decay of the pump laser.}
\end{figure}

$P_3\{E^+ | E^-,E_x\}$ is shown in Fig.~\ref{fig:bispec_srs_2d}. The diagonal streak indicates Stokes decay of the pump: $\omega_1+\omega_2=\omega_0=6.32\omega_p$. SRBS is clearly dominant, and weaker EATS and scattering off BAM noise for $\omega_1<\omega_{1s}$ are also present. Figure \ref{fig:bispec_srs_1d} displays $P_3$ and $b_3$ along the line $\omega_1+\omega_2=6.32\omega_p$; EATS is seen in both as a pair of peaks below 6$\omega_p$. Given the incoherent, Thomson-like nature of EATS, it is not surprising that it is much less bicoherent than SRBS.

To examine BAD, we consider the bispectrum of three spatial Fourier amplitudes $E_x(k,t)$, computed from $E_x(x,t)$. To distinguish left- from right-moving waves, we perform a 2D Fourier transform to the $(k,\omega)$ domain, set the result to zero for all $\omega<0$, and then invert the temporal transform. This procedure can be considered as a Fourier transform in $x$ and a Hilbert transform in $t$. For a wave $E_x=\cos(k_ex-\omega_et)$ this gives $E_x(k_e,t)\propto\exp i\omega_et$. We use $E_{xi} \equiv \mathrm{Re}\ E_x(k_i)$ to compute $P_3$, which requires real fields.

\begin{figure}
\includegraphics{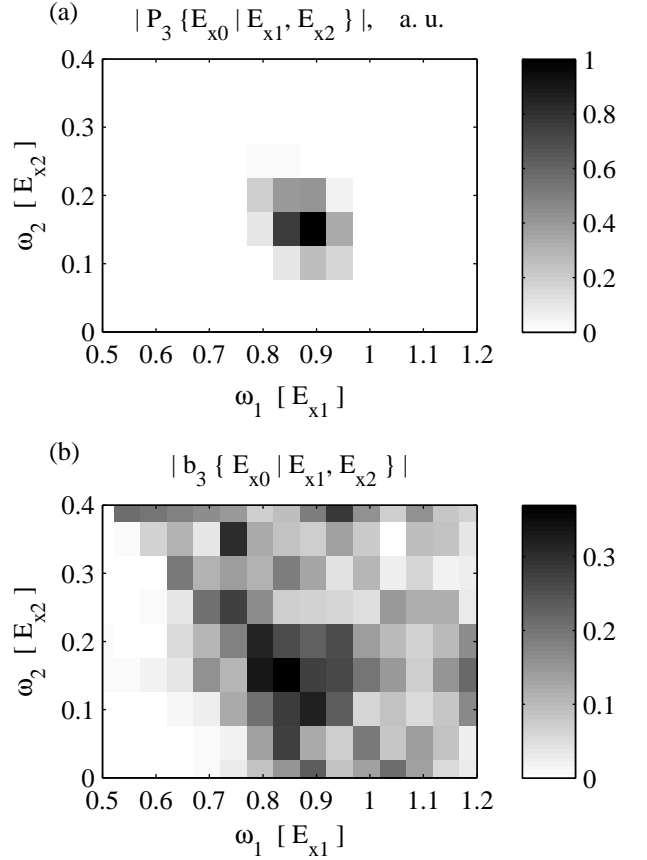}
\caption{(a) Bispectrum and (b) bicoherence of $E_x(k,t)$ Fourier modes $k_i\lambda_D=$ 0.344, 0.209, and 0.136 for $i=0,1,2$, over $t=5-10$ ps in the $I_0=2$ \pw\ Trident run. $E_{xi}=\mathrm{Re}E_x(k_i,t)$. The peak in $|P_3|$ and $|b_3|$ shows parametric coupling in the BAD process.}
\label{fig:bispec_bad}
\end{figure}

Figure \ref{fig:bispec_bad} presents $P_3\{E_{x0}|E_{x1},E_{x2}\}$ and the corresponding $b_3$, where $k_0=k_1+k_2$. The time window of 5-10 ps is when EAW activity is strongest, although similar results obtain for 0-5 ps using slightly different $k$'s. $k_2\lambda_D=0.13$ is chosen slightly below the maximum EAW power, and $k_0\lambda_D=0.344$ gives a larger $P_3$ and more well-defined peak in $b_3$ than do $k_0$'s closer to the SRBS peak of $0.362$. The strongly localized peak in $P_3$ indicates beam acoustic decay, and persists (although at lower absolute amplitude) for nearby $k$ choices. A bicoherence peak, however, is seen only for certain $k$'s, which shows when the EAW is driven by the beating of the BAMs as opposed to satisfying the sum rule with less dynamical coupling. The frequencies $(\omega_0,\omega_1,\omega_2)$ where $P_3$ peaks give $(k,\omega)$ pairs that lie, respectively, slightly below the SRBS peak, near the ``knee'' in the BAM feature, and near the maximum EAW amplitude.

\section{Hohlraum Parameters}
\label{sec:hohl}

\begin{figure}
 \includegraphics{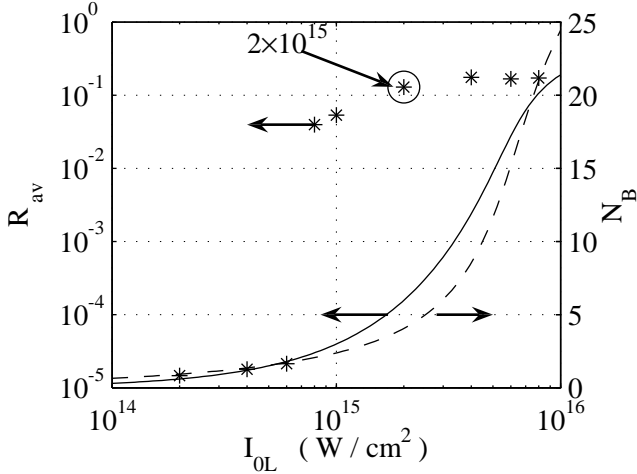}
 \caption{Same as Fig.~\ref{fig:trid_refavg} but for the hohlraum conditions described in the text.}
 \label{fig:hohl_refavg}
 \end{figure}

To demonstrate that kinetically-enhanced SRBS, BAMs, EAWs, and EATS are not peculiar to a narrow parameter range, we show in this section that they occur in ignition hohlraum fill plasmas as well as the single hot spot conditions studied above. We choose a pump wavelength $\lambda_0=(1054/3)$ nm, background density $n_0=0.1n_c$, electron temperature $T_e=3$ keV, and central Krook-free flattop length of 75.2 $\mu$m. The SRBS seed light wave has $\lambda_{1s}=574.8$ nm and $I_{1R}=10^{-5}I_0$, which beats with the pump to produce a plasmon with $k_2\lambda_D=0.357$ (similar to the above Trident value). The time-averaged reflectivity \rav\ versus $I_0$ is plotted in Fig.~\ref{fig:hohl_refavg} and shows an enhancement threshold of $I_0=0.8$ \pw, corresponding to $N_B=2.03$ bounce orbits in the convective steady state. Note that here \rav\ for large $I_0$ saturates near 17\%, which is much larger than the Trident value of 1\%.

\begin{figure}
 \includegraphics{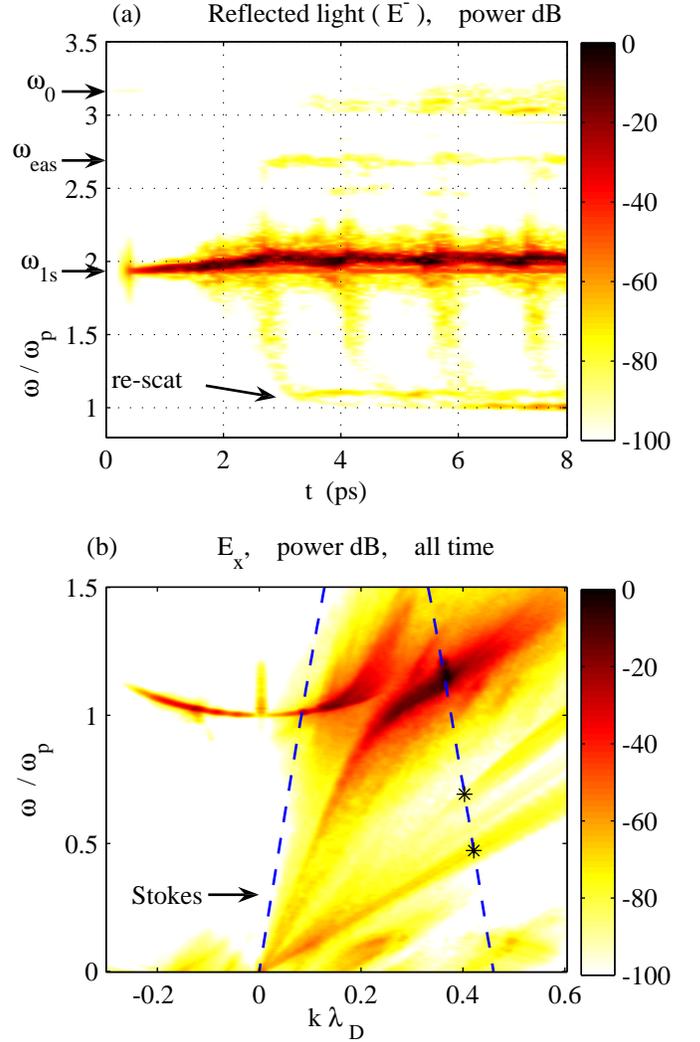}
 \caption{ \label{fig:hohl_emwt_exwk} (Color.) (a) Time-resolved spectrum of reflected light $E^-$ for $I_0=2$ \pw\ hohlraum run. $\omega_{\mathrm{eas}}=2.69\omega_p$ labels the peak of the EAS spectrum, and ``re-scat'' indicates Raman re-scatter of SRBS in its own forward direction. (b) Power spectrum of $E_x$ over the $(x,t)$ domain inside the damping regions. The stars are Stokes points for EAS with scattered light of $\omega/\omega_p$ = 2.47 and 2.69.}
 \end{figure}

We focus on the case $I_0=2$ \pw. Figure \ref{fig:hohl_emwt_exwk}(a) displays the time-resolved reflected light spectrum. As for the Trident runs, the SRBS light upshifts in frequency, at which time EAS develops. The SRBS reflectivity is temporally bursty with large pulses separated by about 1.5 ps. The electrostatic envelope (not shown) is similar to Fig.~\ref{fig:trid_extx}, with four broad pulses propagating away from the laser entrance and correlated to reflectivity peaks. The electrostatic spectrum $E_x(k,\omega)$ is presented in Fig.~\ref{fig:hohl_emwt_exwk}(b). Again, the EPW curve splits into an upper branch and a set of BAMs, and SRBS is downshifted in $\omega$ along the Stokes curve. An EAW is present and mostly energized at $\omega$ much lower than the EAS Stokes point.

As the pump strength $I_0$ is raised, a distinct EAW and EAS persist. This is unlike the Trident runs, where both are washed out by very broad EPW frequency downshifts for strong-pump cases. In addition, reflected light more intense than EAS appears near $\omega_0$ for the hohlraum parameters (visible in Fig.\ \ref{fig:hohl_emwt_exwk}(a)), and becomes stronger with increasing $I_0$. This cannot be stimulated Brillouin scattering since the ion are fixed, although mobile ions may significantly distort this signal. The Stokes curve passes through a broad, low-frequency feature near $k\lambda_D\approx0.45$ in the $E_x$ spectrum, and the $\approx\omega_0$ reflected light may be scattering off this.

\section{Discussion and conclusions}
\label{sec:conc}

Kinetic inflation of SRBS and related physics has been studied with Eulerian Vlasov simulations in single-hot-spot and hohlraum-fill conditions. Inflation of the reflectivity above the coupled-mode steady state level develops suddenly as the pump becomes stronger, in accord with recent Trident experiments. The trapped particle instability may modulate and break up the electrostatic field envelope early in time but does not dominate the subsequent dynamics. Once trapping flattens the distribution, enhanced SRBS occurs in picosecond-scale bursts accompanied by electrostatic pulses propagating away from the laser entrance.

SRBS light upshifts in frequency due to trapping. This can be viewed as due to a nonlinear downshift of the daughter plasmon frequency, following Morales and O'Neil, or as scattering off a lower-frequency linear mode supported by the trapping-modified $f_e$ (in the spirit of quasilinear theory). We adopt the latter approach, and find linear BAMs that agree with the observed electrostatic spectrum and give frequency-downshifted SRBS plasma waves. Our Gauss-Hermite linear analysis reveals a set of BAMs, some of which are unstable without light-wave coupling and thus excitable by purely electrostatic dynamics.

EAWs and EAS have been observed in our simulations and understood in new ways. The Gauss-Hermite method reveals the modified $f_e$ supports a heavily-damped \textit{linear} EAW, which differs from the nonlinear, undamped EAW reported by Rose and others. It is premature to say that EAWs generated by SRBS are a solely linear phenomenon or that the nonlinear theories are irrelevant. Linear explanations nonetheless have a certain appeal (Occam's razor). Future work should explore the impact of de-trapping mechanisms like speckle sideloss, and broadband seeding in Vlasov simulations, on EAW physics. Preliminary results of simulations we have done with a broadband seed show the EAS level is independent of bandwidth.  This supports our EATS interpretation of EAS, as opposed to a stimulated process where seed light in that frequency range would be parametrically amplified.

The connection between our EAS picture and experiment is not yet clear. The relevant Trident experiments (Ref.\ \cite{montgomery-trident-pop-2002}, Fig.\ 10) for $I_0=5$ \pw\ saw EAS light of $\approx 10^{-6}$ the SRBS intensity, similar to the ratio in our run with $I_0=2$ \pw. For pumps above 5 \pw\ the epxeriments see spectrally-narrow and much stronger EAS ($\sim1/3000$ the SRBS intensity), while our simulations for $I_0>2$ \pw\ show no EAS peak but instead strongly-upshifted SRBS and broad, higher-frequency light (as mentioned at the end of Sec.\ \ref{sec:trides}). Moreover, earlier measurements on Trident\cite{cobble-trident-pop-2000} with a random phase plate (RPP) smoothing the interaction beam (giving multiple speckles), which were interpreted in Ref.\ \cite{montgomery-trident-pop-2002} as potentially being SEAS, are much stronger than in our runs. Simulations with different density and temperature (and thus $k_2\lambda_D$) from this paper, but still near the Trident single-hot-spot values, develop EAWs and EAS qualitatively like the results presented here (in particular, the EAS intensity is $\sim10^{-6}$ that of SRBS). Physics not included in this paper's simulations, such as sideloss, ions, or multi-dimensional effects, may provide better experimental contact. Measuring the electrostatic spectrum in the EAW region of the $(k,\omega)$ plane would shed light on EAWs and EAS, in particular revealing if most EAW activity is at lower $k$ and $\omega$ than the EAS matching point (as in our runs).

Our picture of EAW excitation is beam acoustic decay (BAD), or the three-wave interaction of two BAMs and an EAW. Given the very low EAW amplitudes, this is likely a two-pump process, with the daughter BAM primarily generated separately from BAD. The daughter EAWs can weakly couple to higher-$k$ harmonics on the EAW curve. Moreover, $f_e$ is sometimes distorted (but not fully flattened) near the EAW phase velocity; this may lower its damping rate, or facilitate excitation of higher-$k$ EAWs. The pump laser scatters off the higher-$k$ EAW fluctuations, which we think of as electron acoustic Thomson scatter (EATS). We emphasize that the EAWs are strongest well below the matching point for EAS, implying EAS cannot excite them. In addition, the EAW amplitude shows no increase at the EAS matching point, which again suggests that EAS is not significantly energizing the EAWs.

After several decades of work, there are many unanswered questions about SRBS. Simple estimates for the onset and time-averaged reflectivity due to kinetic inflation are being investigated but are not yet established, and the relative importance of the various saturation mechanisms is still unclear. Research on EAWs and EAS, besides its intrinsic physical interest, has value in understanding SRBS: light scattered off EAWs may indicate when distribution functions are strongly modified. Although large EAS reflectivity has yet to be clearly observed, it would not be the first surprise in the rich physics of laser-plasma interactions.

\begin{acknowledgments}     
We thank W.\ M.\ Nevins and C.\ Holland for useful discussions of bispectral analysis, and M.\ M.\ Shoucri for assistance with our Vlasov code.  The work at LLNL was performed under the auspices of the U. S. Department of Energy by University of California, Lawrence Livermore National Laboratory under Contract No.\ W-7405-Eng-48. Part of this work was submitted by D.\ J.\ S.\ in partial fulfillment of the requirements for the degree of Ph.\ D.\ \cite{strozzi-thesis-2005} and was supported in part by Dept.\ of Energy Grant DE-FG02-91ER54109.  Work by A.\ B.\ was supported in part by the Research Lab for Electronics and Plasma Science and Fusion Center, MIT. 
\end{acknowledgments}     

\appendix
\section{Linear modes via Gauss-Hermite projection} 
The Appendix presents how to find the susceptibility and linear modes for an arbitrary distribution by Gauss-Hermite projection. We specialize to the 1-D electrostatic modes of an electron plasma with fixed ions, and choose units where $\omega_p$=$\lambda_D$=$v_{Te}=1$ for a reference $n_0$ and $T_e$. The linear susceptibility $\chi$ is
 \begin{equation}
 \chi(k,\omega) = -k^{-2} {d\over dv_p} \int_{-\infty}^{\infty} dv\ {f \over v-v_p},
 \end{equation}
where $v_p=\omega/k$ is the wave phase velocity, the integral is taken along the Landau contour, and $f$ is the background distribution with $\int dv\, f = n/n_0$. If we write $f=\sum_if_i$ then $\chi=\sum_i\chi_i$ where $\chi_i$ is $\chi$ for $f_i$. This is convenient for our present work, where $f$ consists of a Maxwellian plus a small correction centered at a nonzero $v$. We use a scaled, shifted velocity $u\equiv (v-v_0)/\delta v$ (and similarly for $u_p$) and put $F(u)=f[v(u)]$. We expand $F(u)=\sum_{n=0}^N F_ng_n(u)$ over the basis $\{g_n\}$ and obtain $\chi(k,\omega)=-(k^2\delta v)^{-1}\chi_u(u_p)$ where $\chi_u(u_p) \equiv \sum F_n\chi_{u,n}(u_p)$ and
 \begin{equation}
 \chi_{u,n}(u_p) \equiv {d\over du_p} \int_{-\infty}^{\infty} du\ {g_n(u) \over u-u_p}.
 \end{equation}

A convenient basis, which gives an inherently localized $F$ and analytically known $\chi_{u,n}$, is the Gauss-Hermite basis
 \begin{equation}
 g_n(u) \equiv {1 \over \pi^{1/4}\sqrt{2^nn!}}H_n(u) e^{-u^2/2}.
\end{equation}
$H_n$ is the order $n$ Hermite polynomial (we follow the notation of \cite{absteg}). The $g_n$'s are the orthonormal quantum harmonic oscillator eigenstates: $\int_{-\infty}^\infty du\ g_n(u)g_m(u)=\delta_{nm}$. The projection weights $F_n$ are given by $F_n=\int_{-\infty}^\infty du\ g_n(u)F(u)$.

The $\chi_{u,n}$'s satisfy a recurrence relation for $n\geq2$, found by utilizing properties of $H_n$:
 \begin{equation}
 \label{eq:chirec}
 \chi_{u,n}(u_p) = -\left({2\over n}\right)^{1/2}\chi_{u,n-1}'(u_p) + \left({n-1\over n}\right)^{1/2}\chi_{u,n-2}(u_p).
 \end{equation}
The base cases $n=0,1$ are
 \begin{equation}
 \chi_{u,0}(u_p) = {\pi^{1/4}\over\sqrt2}Z'(w_p); \qquad \chi_{u,1} = -{d\chi_{u,0}\over dw_p}.
 \end{equation}
$w_p=u_p/\sqrt2$, $Z'(w)=dZ/dw$, and $Z$ is the plasma dispersion function \cite{fried-zfunc-1961}:
 \begin{equation}
 Z(w_p) \equiv \pi^{-1/2} \int_{-\infty}^{\infty} dw\ {e^{-w^2} \over w-w_p}.
 \end{equation}
 We evaluate $Z$ with the numerical algorithm in Ref.~\cite{weideman-cef-siamjna-1994}, which provides a high-order rational function approximation valid throughout the complex plane. The derivatives of $Z$ are a polynomial times $Z$ plus a remainder polynomial. Therefore, we have 
\begin{equation}
 \label{eq:chipoly}
 \chi_{u,n}(u_p) = K_{Z,n+1}(u_p)Z(w_p) + K_{R,n}(u_p)
\end{equation}
where $K_{Z,n}$ and $K_{R,n}$ are order $n$ polynomials. Eq.~(\ref{eq:chirec}) yields a recurrence relation among the polynomials:
 \begin{eqnarray}
 K_{Z,n+1}(u) &=& \left( 2\over n \right)^{1/2} u K_{Z,n} - 
                 \left( 2\over n \right)^{1/2} K_{Z,n}' \\
              && + \left( n-1 \over n \right)^{1/2} K_{Z,n-1},  \\
 K_{R,n}(u) &=& {2\over\sqrt n}K_{Z,n} - 
             \left(2\over n\right)^{1/2}K_{R,n-1}' \\
           && + \left(n-1\over n\right)^{1/2}K_{R,n-2}
 \end{eqnarray}
These formulas hold for $n\geq2$ and allow the polynomial coefficients to be pre-computed. The base cases are $K_{R,0}(u)=-\pi^{1/4}\sqrt2$, $K_{R,1}=-2\pi^{1/4}u$, $K_{Z,1}=-\pi^{1/4}u$, and $K_{Z,2}=\pi^{1/4}\sqrt2(1-u^2)$.

 The linear modes are the roots of the dispersion relation $\epsilon\equiv 1+\chi(k,\omega)=0$. We find complex $\omega$ for real $k$ using Newton's method. This requires $\chi_{u,n}'(u_p)$, which from Eq.~(\ref{eq:chipoly}) for $n\geq2$ is
 \begin{equation}
 \chi_{u,n}'(u_p) = L_{Z,n+2}(u_p)Z(w_p) + L_{R,n+1}(u_p).
 \end{equation}
$L_{Z,n}$ and $L_{R,n}$ are order $n$ polynomials given by (again for $n\geq2$)
 \begin{eqnarray}
 L_{Z,n+2}(u) &=& -uK_{Z,n+1}(u) + K'_{Z,n+1}(u), \\
 L_{R,n+1}(u) &=& -\sqrt2 K_{Z,n+1}(u) + K_{R,n}'(u).
 \end{eqnarray}

The Gauss-Hermite method has several advantages over other approaches. Numerically performing the Landau integral for every $v_p$ of interest is time-consuming, requires care in handling the Landau contour, and needs $f$ to be analytically continued off the real $v_p$ axis. Our technique only computes several integrals for the $F_n$'s, automatically handles the Landau contour via the $Z$ function, and is valid throughout the complex plane. Writing $f$ as a series of Maxwellians involves some guesswork since Maxwellians do not form an orthogonal basis. One can use a different scale for $H_n$ and the exponential, for instance $\hat{g}_n(u)\propto H_n(u)e^{-u^2}$. This gives a simpler recurrence relation for $\chi_{u,n}$, but requires more terms to reconstruct $f$ well than the $g_n$ basis does. We have tried both bases and find similar linear modes. 

Unfortunately, the modes given by any Gauss-Hermite basis do not converge as $N\rightarrow\infty$. For $u$ far inside the classical turning points (in analogy with the quantum oscillator), $g_n\sim\sin(\kappa_nu+\phi_n)$ where $\kappa_n\sim\sqrt{2n+1}$, as a WKB analysis easily shows. Higher-order $g_n$'s thus involve oscillation on smaller velocity scales. When extended to complex $u$, these oscillations give rise to exponential growth. The $\chi_{u,n}$ similarly blow up in the lower half-plane (as is well-known for the $Z$ function), with $|\chi_{u,n}|$ growing with $n$ for fixed $u$ below the real axis. Very small-scale features in $f$ project onto high-$n$ basis functions $g_n$, which produce rapidly growing $\chi_{u,n}$ in the complex plane and unphysically distort the modes. As $N$ increases for the parameters of Fig.~\ref{fig:trid_chi}, the roots vary slightly and a single zero-contour of $\epsilon$ can bifurcate into two. Low-pass filtering before projection could mitigate this. Although the Gauss-Hermite method gives much insight into the linear dynamics of an arbitrary $f$, its use requires caution and practice.




\end{document}